\documentclass[superscriptaddress,amsmath,amssymb,aps,prb,onecolumn]{revtex4-2}

\usepackage{graphicx}
\usepackage{dcolumn}
\usepackage{bm}
\usepackage[utf8]{inputenc}
\usepackage[T1]{fontenc}
\usepackage{mathptmx}
\usepackage{afterpage}

\usepackage{color}
\usepackage{soul}
\newcommand{\RNum}[1]{\uppercase\expandafter{\romannumeral #1\relax}}
\usepackage[super,negative]{nth}
\usepackage{xcolor}
\usepackage{hyperref}

\begin{document}

\title{Topological phase transition in disordered elastic quantum spin Hall system}

\author{Xiaotian Shi}
\affiliation{Department of Aeronautics and Astronautics, University of Washington, Seattle, Washington 98195, USA}%

\author{Rajesh Chaunsali}
\affiliation{Department of Aerospace Engineering, Indian Institute of Science, Bangalore 560012, India}%
   
\author{Georgios Theocharis}
\affiliation{LAUM, CNRS, Le Mans Université, Avenue Olivier Messiaen, 72085 Le Mans, France}%

\author{Huaquing Huang}
\thanks{huanghq07@pku.edu.cn}
\affiliation{School of Physics, Peking University, Beijing 100871, China.}%

\author{Rui Zhu}
\thanks{ruizhu@bit.edu.cn}
\affiliation{School of Aerospace Engineering, Beijing Institute of Technology, Beijing 100081, China}%

\author{Jinkyu Yang}
\thanks{jkyang@aa.washington.edu}
\affiliation{Department of Aeronautics and Astronautics, University of Washington, Seattle, Washington 98195, USA}
\affiliation{Department of Mechanical Engineering, Seoul National University, Seoul, 08826, Republic of Korea}%


\begin{abstract}
We investigate the effect of disorder on topologically nontrivial states in a two dimension (2D) mechanical system. We first propose a quantum spin Hall (QSH) insulator based on an out-of-plane spring-mass model and analytically study the interplay between the disorder and topology in both topologically trivial and nontrivial systems. We adopt the spin Bott index to characterize the topological property in disordered mechanical systems. By tracking the evolution of the spin Bott index with the increase of disorders, we quantitatively demonstrate the disorder induced transition from a topologically nontrivial QSH insulator to a trivial insulator. We then validate the topological phase transition through transient analysis in discrete lattices. Finally, we design a phononic crystal based on the discrete spring-mass model and numerically verify the topologically protected states along the boundary between the trivial insulator and disordered topological QSH insulator in a continuous system. This work puts a step forward in understanding the role of disorder in a 2D topological classical system.

\end{abstract}

\maketitle

\section{Introduction}
The discovery of topological insulators (TI)~\cite{qi2011,hasan2010} opened a new era for studies in condensed matter physics. The key feature of a TI is the robust and directional flow of energy at the boundary of the system, which promises a powerful tool for the design of low-dissipative devices. In general, the topological insulators can be classified into two kinds, those that break the time-reversal (TR) symmetry (e.g., the quantum Hall (QH) insulator~\cite{klitzing1980,thouless1982}) and those that preserve the time-reversal symmetry (e.g., the quantum spin Hall (QSH) insulator~\cite{kane2005,bernevig2006} and the quantum valley Hall (QVH) insulator~\cite{rycerz2007,xiao2007}). As the former type usually requires additional efforts such as the strong external magnetic fields to support topological states, the TR-symmetry preserved system is more practical to realize, thus drawing significant attention recently. 

The QSH effect can be regarded as the effect of two coupled quantum Hall states with opposite Chern numbers for each spin. Taking advantage of the spin-orbit coupling~\cite{bernevig2006}, the QSH system can support robust helical edge modes even in the absence of magnetic fields. However, due to the lack of intrinsic spin, it is not straightforward to realize the quantum spin hall phase in classic wave systems which are composed of spin-less particles. In order to generate such a pseudospin degree of freedom, Wu \textit{et al}. introduced the zone-folding technique and reported a topological photonic crystal with $C_6$ lattice symmetry~\cite{wu2015}. Afterward, similar ideas have been applied to acoustic~\cite{he2016} and mechanical~\cite{yu2018} wave systems, which helped to take a great step in the development of topological metamaterials.

One of the important research directions in the study of topological insulators is to understand the effect of disorder on topology~\cite{hasan2010,qi2011}. Generally, the topological state is immune to weak disorder. With the increase of the disorder level in the system, the topologically nontrivial phase will eventually vanish. Such topological phase transition is intuitive to understand since the Bloch theory fails in the presence of disorder that suppresses the periodicity of the structure. Moreover, further investigations found an abnormal transition from a trivial system to a topological one solely induced by disorder, accompanied by the realization of topological Anderson insulator (TAI)~\cite{li2009a}. Followed by the early works on phase transitions in disordered topological electronic systems~\cite{jiang2009a, groth2009,bardarson2010,guo2010,nomura2011,schubert2012,song2012,kobayashi2013,mondragon-shem2014,song2014}, the study on the interplay between disorder and topology also extends to other fields, such as photonic systems~\cite{liu2017,stutzer2018,yang2020,liu2020,zhang2022} and electric circuits~\cite{zhang2019,zhang2021}. Research on disorder induced topological phase transition in the elastic topological system has also been reported in one dimension (1D) system. Zangeneh-Nejad.~\textit{et al}. realized the topological Anderson insulator phases in a topological sonic crystal~\cite{zangeneh-nejad2020a}. Shi.~\textit{et al}. numerically investigated the topological phase transition in a disordered spring-mass chain~\cite{shi2021}. 

Unlike the 1D systems, qualitative analysis and verification of the topological transition in two dimension (2D) elastic systems is still challenging due to the following reasons: (1) Lack of proper topological invariant in a 2D disordered system. Without explicit band structure, traditional classic topological invariants, such as the Chern number and spin Chern number, are ill-defined. (2) Difficulty in precisely manipulating the disorder in 2D topological metamaterials. Particularly, accurate mapping from disorder parameters in analytical models to a structure is not easy to achieve. Recently, Liu \textit{et al}. reported observation of TAI in a 2D acoustic spin Chern insulator~\cite{liu2021}. They proposed a bilayer phononic crystal with synthetic spin-orbit coupling and experimentally demonstrated disorder-driven topological spin-dependent edge states. However, disorder induced topological phase transition in other types of 2D elastic TI is still elusive. Particularly, analysis of the disorder effect in zone folding induced TI, which is the most common type of classical analogy of TI, is yet to be explored.

In this work, we focus on the investigation of disorder induced topological phase transition in 2D elastic systems. First, we start with a discrete mechanical QSH insulator based on an out-of-plane spring-mass lattice. We compare different approaches to calculate the topological invariant in periodic cases and then  generalize the spin Bott index to work as an effective spin Chern number to characteristic the topology of disordered elastic QSH systems. Based on that, we systematically analyze the effect of random stiffness disorder on the topology and quantitatively demonstrate the disorder induced topological phase transition. Then, we validate the analysis by comparing the transient wave fields of finite size disordered structures under pseudospin excitation. Finally, we propose a phononic crystal and verify computationally the disorder induced topological phase transition process in continuous systems.

\section{Mechanical quantum spin Hall insulator} \label{C5S2}

\begin{figure}[h!]
\includegraphics[width=0.8\columnwidth]{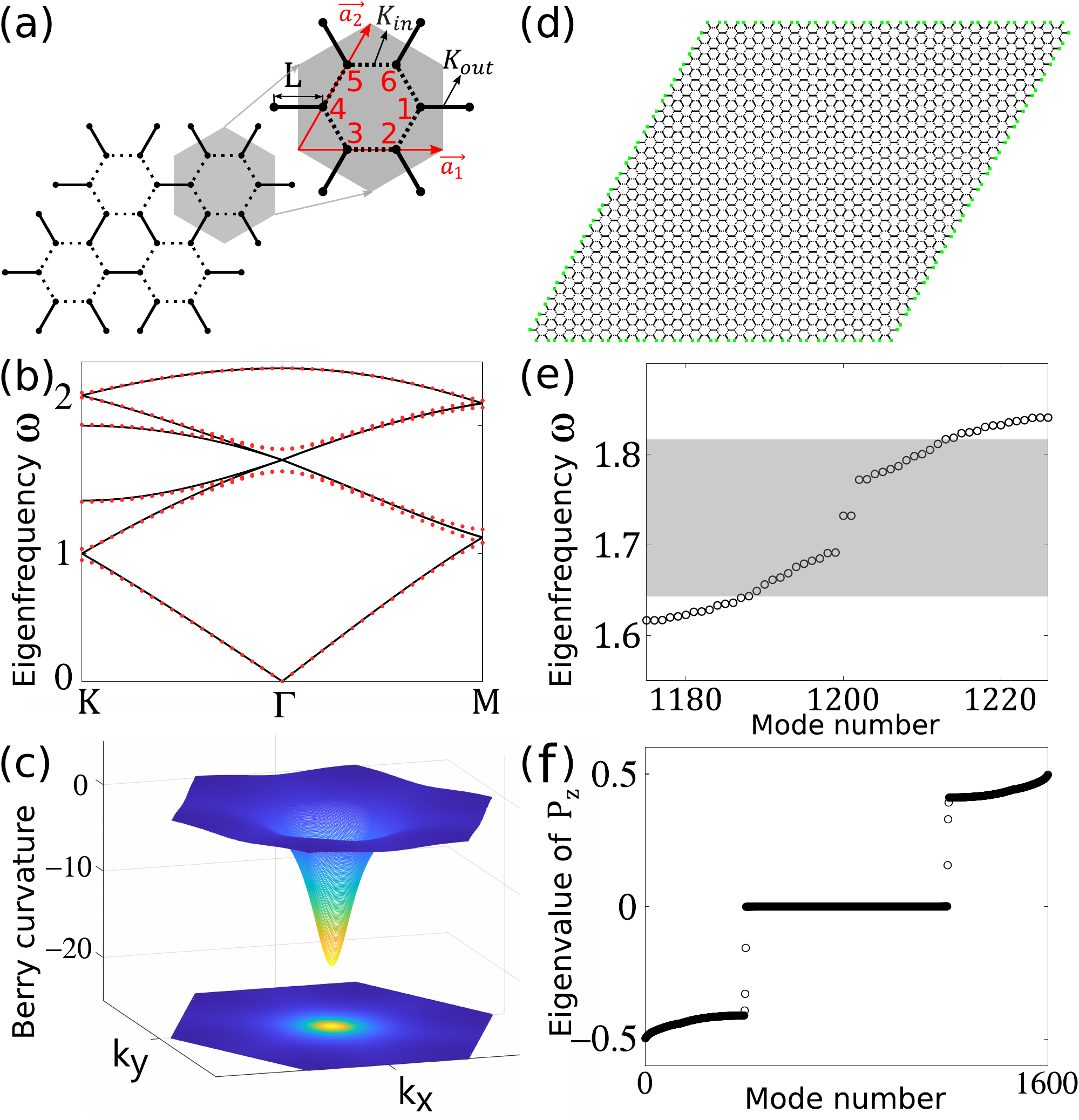}
\caption{(a) A schematic of the unit cell of a spring-mass-based discrete structure. (b) Band structure of the expanded unit cell. (c) Berry curvature of the lower band of the spin-up channel. (d) Configuration of a finite size lattice structure with fixed boundary conditions. Green dots stand for the mass particles whose displacements are set to zero. (e) Eigenfrequencies of the finite size structure. (f) Eigenvalues of the spin projector operator $P_z$.}
\label{C5FIG1}
\end{figure}

We start with a simple mechanical QSH model, which is composed of particles connected by springs. As shown in Fig.~\ref{C5FIG1}(a), the system is a honeycomb lattice with uniform masses ($m=1$).  Following the zone-folding mechanism~\cite{wu2015}, we choose an expanded unit cell containing $6$ particles as indicated by the shaded hexagon. The distance between the nearest neighboring masses is equal to $L$. $\vec{a_{1}} = a(1,0) $ and $\vec{a_{2}} = a(\frac{1}{2}, \frac{\sqrt{3}}{2})$ are lattice vectors with $a=3L$ being the lattice constant, as indicated by the red arrows in Fig.~\ref{C5FIG1}(a). 

We then define springs connecting masses within a unit cell as intracell springs with stiffness $K_{in}$, shown as the dashed black lines in Fig.~\ref{C5FIG1}(a). The springs linking masses of different cells are noted as intercell springs with stiffness $K_{out}$, shown as solid black lines in Fig.~\ref{C5FIG1}(a). Note that we will focus on the out-of-plane motion of the mass particles only. That is, each particle has just one degree of freedom. By applying Floquet Bloch boundary conditions, we can get the dynamic matrix of the unit cell given as:
\begin{equation}
\textbf{\textit{D}} = \frac{1}{m} \begin{bmatrix}
 K_{d}&  -K_{in}&  0&  -K_{out}e^{i\vec{k}\vec{a_{1}}}&  0&-K_{in} \\ 
 -K_{in}&  K_{d}&  -K_{in}&   0&  -K_{out}e^{i\vec{k}(\vec{a_{1}}-\vec{a_{2}})}& 0\\ 
 0&  -K_{in}&  K_{d}&  -K_{in}& 0 & -K_{out}e^{-i\vec{k}\vec{a_{2}}}\\ 
 -K_{out}e^{-i\vec{k}\vec{a_{1}}}&  0&  -K_{in}&  K_{d}&  -K_{in}& 0\\ 
 0&  -K_{out}e^{-i\vec{k}(\vec{a_{1}}-\vec{a_{2}})}&  0&  -K_{in}&  K_{d}& -K_{in}\\ 
 -K_{in}&  0&  -K_{out}e^{-i\vec{k}\vec{a_{2}}}&  0&  -K_{in}& K_{d}
\end{bmatrix}
\end{equation}

\noindent where $K_{d}=2K_{in}+K_{out}$, $\vec{k}=(k_{x},k_{y})$, and $k_{x}(k_{y})$ is the wave number in $x(y)$ direction. 

The dispersion relation can be obtained by solving the eigenvalue problem given as: 

\begin{equation}
    \omega^2 \textbf{\textit{D}} \textbf{\textit{U}} = \textbf{\textit{U}} 
\end{equation}

\noindent where $\omega$ is the angular frequency and $\textbf{\textit{U}}= [u_{1}, u_{2}, u_{3}, u_{4}, u_{5}, u_{6}]^{T}$ is the displacement vector constituted from the nodal displacement for each particle within the unit cell. Figure~\ref{C5FIG1}(b) shows the dispersion curves of the unit cell with different arrangements of $K_{in}$ and $K_{out}$. The black lines represent the case for $K_{in}=K_{out}=K_{0}=1$, where a four-band degeneracy can be observed at $\Gamma$ point ($\vec{k}=(k_{x},k_{y})=(0,0)$). When $K_{in}$ and $K_{out}$ are different, the degeneracy will be lifted and a band gap appears. To ensure that the band gap is formed near the original double Dirac point, we set the intracell and intercell spring stiffness as $K_{in} = K_{0}(1-0.5\Delta)$ and $K_{out} = K_{0}(1+\Delta)$. Here, $\Delta$ represents the perturbation of the spring stiffness, and it should be noted that the levels of perturbation are different between $K_{in}$ and $K_{out}$ to ensure the band gap is formed near the original Dirac frequency. By varying the relative stiffness of intercell and intracell springs, the system can be switched between a trivial insulator and a QSH insulator. The red dotted lines in Fig.~\ref{C5FIG1}(b) stand for a case for $K_{in}=0.9<K_{out}=1.2$ with a band gap being formed near the original double Dirac point. 

To characterize its topological property, we can calculate a topological invariant associated with the system, such as, the spin Chen number $C_{s}$. One way to determine the spin Chern number is through the analytical approach, where we need to derive the effective Hamiltonian of the system and map it to the Bernevig-Hughes-Zhang (BHZ) model~\cite{bernevig2006,wu2016} of QSH effect in an electron system. 
By transforming the dynamical matrix $\textbf{\textit{D}}$ to a pseudospin vector basis \(\begin{bmatrix}
p_+, & d_+, & p_-, & d_-
\end{bmatrix}\), we can obtain the effective Hamiltonian of the system, given as:
\begin{equation}
    \begin{split}
        \textbf{\textit{H}} & = {Q}^{\dagger}{[ p_x, d_{xy}, p_y, d_{x^2-y^2}]}^{\dagger} \textbf{\textit{D}} [ p_x, d_{xy}, p_y, d_{x^2-y^2}]Q \\
          & = {[ p_+, d_+, p_-, d_-]}^{\dagger} \textbf{\textit{D}} [ p_+, d_+, p_-, d_-] \\
          & = {T_{s2n}}^{\dagger} \textbf{\textit{D}} T_{s2n}
    \end{split}
\end{equation}

\noindent where \begin{math}
\begin{bmatrix}
p_x, & d_{xy}, & p_y, & d_{x^2-y^2}
\end{bmatrix}
= 
 \begin{pmatrix}
   \frac{\sqrt{3}}{3} & \frac{\sqrt{3}}{6} & -\frac{\sqrt{3}}{6} & -\frac{\sqrt{3}}{3} & -\frac{\sqrt{3}}{6} & \frac{\sqrt{3}}{6} \\
  0 & -\frac{1}{2} & \frac{1}{2} & 0 & -\frac{1}{2} & \frac{1}{2} \\
  0 & -\frac{1}{2} & -\frac{1}{2} & 0 & \frac{1}{2} & \frac{1}{2} \\
  -\frac{\sqrt{3}}{3} & \frac{\sqrt{3}}{6} & \frac{\sqrt{3}}{6} & -\frac{\sqrt{3}}{3} & \frac{\sqrt{3}}{6} & \frac{\sqrt{3}}{6} \\

 \end{pmatrix} ^{T}
 \end{math} 
 represent the electronic orbital-like $p/d$ type degenerate modes at the $\Gamma$ point, 
 \begin{math} Q=  \begin{pmatrix}
  \frac{1}{\sqrt{2}} & 0 & \frac{1}{\sqrt{2}} & 0 \\
  0 & \frac{1}{\sqrt{2}} & 0 & \frac{1}{\sqrt{2}} \\
  \frac{i}{\sqrt{2}} & 0 & \frac{-i}{\sqrt{2}} & 0 \\
  0 & \frac{i}{\sqrt{2}} & 0 & \frac{-i}{\sqrt{2}} 
 \end{pmatrix}\end{math} is the unitary transformation operator, and \begin{math} T_{s2n} = \begin{bmatrix}
p_+, & d_+, & p_-, & d_-
\end{bmatrix}\end{math} is the transformation matrix from natural basis to spin vector basis, which is constituted by the pseudospin up ($+$) and down($-$) states.
Specifically, $p_{\pm} = \frac{1}{\sqrt{2}}(p_x \pm ip_y)$ and $d_{\pm} = \frac{1}{\sqrt{2}}(d_{xy} \pm id_{x^2-y^2})$.

Then we need to expand the effective Hamiltonian $\textbf{\textit{H}}$ near $\Gamma$ point to the first order and get:
\begin{equation}
\textbf{\textit{H}}(\delta \vec{k})=\begin{bmatrix}
\textbf{\textit{H}}_{+}(\delta \vec{k}) & 0\\ 
 0& \textbf{\textit{H}}_{-}(\delta \vec{k})
\end{bmatrix} 
\end{equation}

\noindent with 
\begin{equation}
\textbf{\textit{H}}_{\pm}(\delta \vec{k}) = \begin{bmatrix}
K_{d}-(K_{in}-K_{out})+ \frac{1}{2} a^2 {\delta k}^2   &  \frac{1}{2} a K_{out} (\mp \delta k_x -i\delta k_y ) \\ 
\frac{1}{2} K_{out} (\mp \delta k_x +i \delta k_y  & K_{d}+(K_{in}-K_{out})- \frac{1}{2} a^2 K_{out} {\delta k}^2
\end{bmatrix} 
\end{equation}

\noindent where $\vec{k}=(\delta k_x, \delta k_y)$ is a small wave vector deviating from the $\Gamma$ point, ${\delta k}^2 = {\delta k_x}^2 + {\delta k_y}^2$, and $\textbf{\textit{H}}_{\pm}(\delta \vec{k})$ represents the Hamiltonian for the pseudospin-up/down state. Note that, $\textbf{\textit{H}}(\delta \vec{k})$ resembles the BHZ Hamiltonian of HgTe quantum wells~\cite{bernevig2006}.

The other way to obtain the spin Chern number is to numerically integrate the Berry curvature of  band dispersion of the projected effective Hamiltonian $\textbf{\textit{H}}_{\pm} (\delta \vec{k})$ over the first Brillouin zone. Figure~\ref{C5FIG1}(c) depicts the Berry curvature ($\Omega$) of the lower spin-down channel for $K_{in}<K_{out}$. By integrating the Berry curvature over the first Brillouin zone, we can obtain the corresponding spin Chern number given as $ C_{s}^{-\downarrow} = \frac{1}{2\pi} \int_{BZ} \Omega  \mathrm{d}^2 k $. See Appendix A for the Berry curvatures and spin Chern numbers of all spin channels. Following these methods, we can find that the system is a trivial insulator ($ C_{s} = 0 $) for $K_{in}>K_{out}$ or a nontrivial QSH insulator ($ C_{s} = 1 $) for $K_{in}<K_{out}$. By combining these two kinds of topologically distinct phases, we can realize spin-dependent edge states along the domain wall. Here, the spin Chern number provides an elegant way to characterize the topological property in the proposed periodic system. 

\section{Spin Bott index in Mechanical systems}

In the presence of disorder, the translational symmetry of the structure is broken. The two aforementioned classic methods to calculate the spin Chern number in the periodic system are not applicable anymore. Therefore, we need to use a new topological invariant to classify the topological phases in disordered QSH systems. Here, we adopt the spin Bott index proposed by Huang \textit{et al.}~\cite{huang2018,huang2018a} in electronic systems and generalize it to a mechanical setup. In this work, we focus mainly on spring stiffness disorder. Note that our spring-mass model can be treated as a special phononic 2D Su-Schrieffer-Heeger (SSH) model in the study of the disorder~\cite{coutant2020}. Specifically, the disorder in the spring stiffness simultaneously perturbs the hopping strength and on-site potential in the SSH model. 
We impose a global random stiffness disorder in the system, given as:
\begin{equation} \label{C5eq:5}
    \begin{aligned} 
    k_j &= K_j + K_{0}W_j \epsilon_j & \text{ for } j &= 1,2,...,N \\
\end{aligned}  
\end{equation}

\noindent where $N$ is the total number of springs, $\epsilon_j$ is a random number ranging from $-1$ to $1$, $K_j$ is the unperturbed spring stiffness and may take the values $K_{in}$ or $K_{out}$ depending on its location, $W_j= W\xi_j$ is the corresponding disorder magnitude with $W$ being the disorder strength, and $\xi_j$ is a coefficient that controls the ratio between the disorder strength on intracell and intercell springs. 

To calculate the spin Bott index, we consider a finite size structure containing $N=N_{1} \times N_{2}$ unit cells and impose fixed boundary conditions on all four boundaries as shown in Fig.~\ref{C5FIG1}(d). The dynamical matrix of the finite size structure is noted as $\mathbb{D}$. Throughout our study on discrete models, we set $N_{1}=N_{2}=20$. First, we need to derive the effective Hamiltonian $\mathbb{H}$ of this finite size structure. See Appendix B for more details about the construction of $\mathbb{D}$ and $\mathbb{H}$.

Then, we can construct a projector operator of the occupied states of the effective Hamiltonian $\mathbb{H}$ below the band gap,
\begin{equation}
P = \sum_{i}^{N_{occ}} | \varphi_i \rangle \langle \varphi_i |,
\end{equation}
\noindent where $N_{occ} = 2N$, $\varphi_i$ is the $i$th normalized eigenvector of the effective Hamiltonian $\mathbb{H}$ corresponding to eigenvalue $\omega_i^2$. Figure~\ref{C5FIG1}(e) shows the eigenfrequencies of the finite size clean structure (without disorder $W=0$) made of the unit cell defined in Fig.~\ref{C5FIG1}(a) with $K_{in}=0.9 < K_{out}=1.2$. The shaded region represents the band gap predicted by the unit cell band structure. Please find the eigenfrequencies and eigenmodes of the finite size structure with different types of unit cells in Appendix C. 

Next, we need to decompose $P$ into two spin sectors by using a spin operator $\widehat{S_{z}}$
\begin{equation}
P_{z} = P \widehat{S_{z}}P,
\end{equation}

\noindent where $\widehat{S_{z}} = S_{z} \bigotimes I_N$, and $S_{z} = \left[\begin{smallmatrix}1 & 0 & 0 & 0 \\ 0 & 1 & 0 & 0 \\ 0 & 0 & -1 & 0 \\ 0 & 0 & 0 & -1\end{smallmatrix}\right]$. Figure~\ref{C5FIG1}(f) shows the eigenvalues of $P_z$. One can see nonzero eigenvalues of $P_z$ are generally divided into two groups ($| \pm \phi_i \rangle = \pm \frac{1}{2}$) corresponding to the spin up and spin down sectors, respectively. We can write the associated eigenvalue problem as:
\begin{equation}
P_{z}| \pm \phi_i \rangle = S_{\pm} | \pm \phi_i \rangle.
\end{equation}

\noindent Accordingly, the new spin up/down $(+/-)$ projector operator can be expressed as:
\begin{equation}
P_{\pm}| = \sum_{i}^{N_{occ}/2} | \pm \phi_i \rangle \langle \pm \phi_i |.
\end{equation}

After rescaling the coordinate of the centroid of each unit cell into the interval $[0,1)$, one can derive the projected position operators as:
\begin{align}
U_{\pm}| = P_{\pm}e^{i2 \pi X}P_{\pm} + (I - P_{\pm}),\\
V_{\pm}| = P_{\pm}e^{i2 \pi Y}P_{\pm} + (I - P_{\pm}),
\end{align}

\noindent Finally, we can calculate the Bott index of the two spin sectors. The spin Bott index is given as~\cite{huang2018,huang2018a}:
\begin{align}
B_{\pm} &= \frac{1}{2 \pi} Im \{tr[log( V_{\pm} U_{\pm} V_{\pm}^{\dagger} U_{\pm}^{\dagger} )] \},\\
B_s &= \frac{1}{2}(B_+ - B_-).
\end{align}

\noindent Note that the spin Bott index is equivalent to the spin Chern number and can be applied to disordered systems to determine the topological QSH phases. With such a powerful tool, we can start to explore topological transitions in the disordered 2D mechanical lattice.

\section{Disorder induced topological transition in discrete mechanical lattice} \label{C5S4}

\begin{figure}[ht]
\includegraphics[width=\columnwidth]{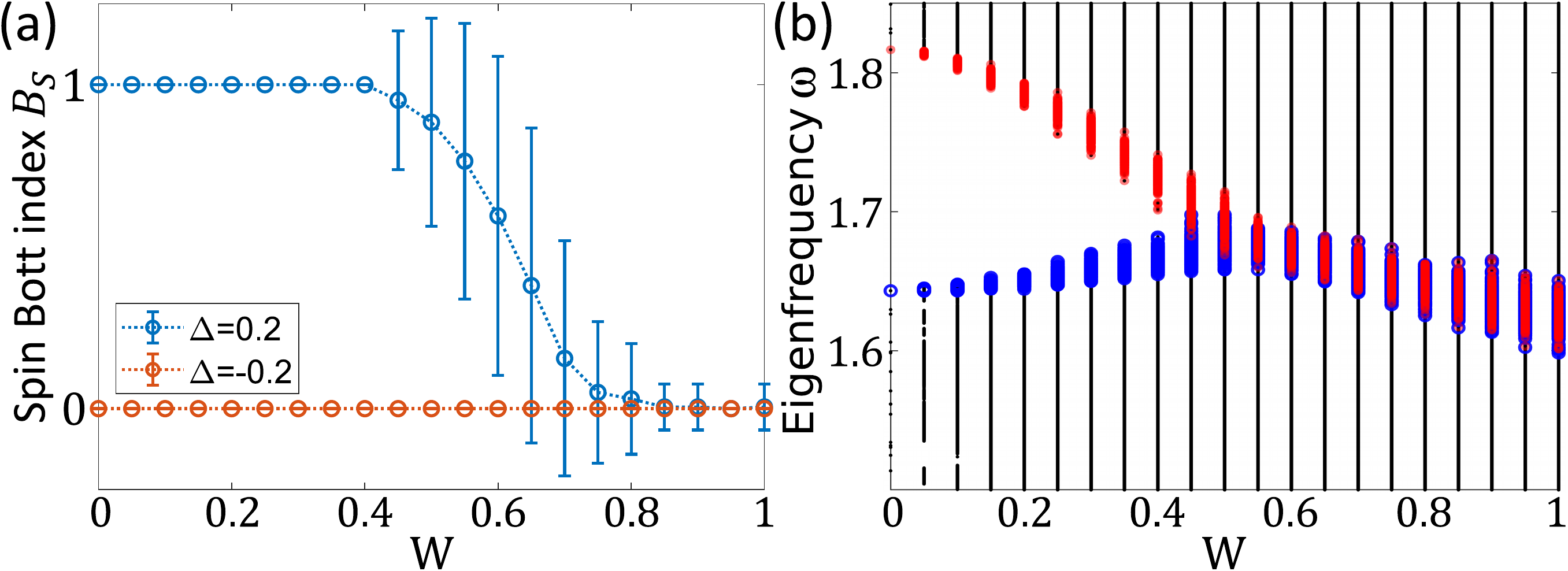}
\caption{(a) Dependence of spin Bott index $B_{S}$ on the disorder strength under random discrete disorder. (b) Eigenfrequencies of a topologically nontrivial structure with periodic boundary conditions under different levels of disorder.}
\label{C5FIG2}
\end{figure}

To demonstrate the disorder induced topological transition, we take the structure shown in Fig.~\ref{C5FIG1}(d) with $20$ unit cells in both lattice vector directions. The finite structure contains $2400$ particles thus $2400$ degrees of freedom in total. In this study, we propose a special type of disorder, a \textit{\textbf{discrete disorder}}, and explore its effects on topology in both topologically trivial and nontrivial $2$D mechanical lattice systems. Specifically, the discrete disorder is determined by the choice of random parameter $\epsilon_j$ in Eq.~\ref{C5eq:5}. Here $\epsilon_j$ is drawn from the discrete uniform distribution on the interval $[-1,1]$ and can only take values $-1$, $0$, or $1$. Following the ratio of perturbation in defining $K_{in}$ and $K_{out}$, we set $\xi_{in}=0.5$ and $\xi_{out}=1$. We choose $K_{0}=1$ and $\Delta = \mp 0.2$ to define our clean systems (no disorder) as the base reference. Based on our previous analysis of the periodic discrete model in section~\ref{C5S2}, we know that when $\Delta = - 0.2$, the system is topologically trivial with $K_{in}=1.1 > K_{out}=0.8$. For $\Delta = 0.2$, the system becomes topologically nontrivial with $K_{in}=0.9 < K_{out}=1.2$.

Figure~\ref{C5FIG2}(a) shows the evolution of the spin Bott index with the increase of disorder strength $W$. Each data point represents the numerical average of $200$ disorder realizations and the error bars indicate the standard deviations. We begin by taking a look at the role of disorder in a trivial system. As shown by the red line in Fig.~\ref{C5FIG2}(a), the spin Bott index of a trivial system remains zero regardless of the disorder strength. That is, introducing random discrete disorder cannot alter topological phases in the trivial mechanical lattice for our choice of $\xi_{in}$ and $\xi_{out}$. Then, we discuss the phase transition in a topologically nontrivial system under disorder, indicated by the blue line in Fig \ref{C5FIG2}(a). For $W=0$, such a clean system is a QSH insulator with the spin Bott index being $1$, which agrees well with our analytical results obtained from the unit cell analysis. After introducing disorders in the system, we can see that the spin Bott index remains constant and barely shows any deviation at the low disorder level. This is clear evidence of the robustness of the topological property which is proven to be immune to weak disorder ($W<0.4$). By further increasing the disorder strength, we observe a sharp drop of the spin Bott index from $1$ which eventually becomes quantized near $0$. This indicates that the system is changing from a topologically nontrivial phase to a trivial one. 

This topological phase transition is usually accompanied by the closure of a band gap. We take the structure shown in Fig.~\ref{C5FIG1}(d) to investigate the band gap closing process. To get rid of the influence of boundary modes, we modify the structure by imposing periodic boundary conditions at all four edges. Figure~\ref{C5FIG2}(b) shows the eigenfrequencies of the 20 by 20 structure with the increase of disorder strength. At each disorder level, we take 200 realizations and plot all the eigenfrequencies overlapping with each other. Blue and red circles are the lower and upper edges of the band gaps for all realizations, which refers to the center two eigenfrequencies (\nth{1200} and \nth{1201}) in the spectrum. One can see that the band gap always exists for all the disorder realizations when the disorder is in a weak range $W<0.4$, which very well explains the initial plateau of spin Bott index under disorder in Fig.~\ref{C5FIG2}(a). 

Upon further increase of disorder, the upper and lower edges of the disordered topological band gap start to merge with each other. Particularly, during the transition region, we see very large deviations in the disorder-averaged spin Bott index. This is because the system's topological nature becomes very sensitive to small perturbations near the critical point and is highly dependent on specific realization when the band gap is about to close. In summary, we quantitatively demonstrate that the disorder in the QSH system tends to destroy its topological nature. Note that the disorder in our model is determined by two parameters, $\xi_j$, and $\epsilon_j$. By varying these two parameters, we can manipulate the topological phase transition process by shifting the critical transition boundary. Please find more details about the effect of disorder parameters in Appendix D.

\section{Pseudospin-dependent interface states in disordered QSH system}
One of the most important features of the QSH system is the unidirectional propagation of topological edge modes. To validate previous analysis on topological phase transition by tracking the spin Bott index, we proceed to investigate the transient pseudospin-dependent wave transport along a designed interface between a trivial insulator and a disordered QSH TI.

\begin{figure}[ht]
\includegraphics[width=\columnwidth]{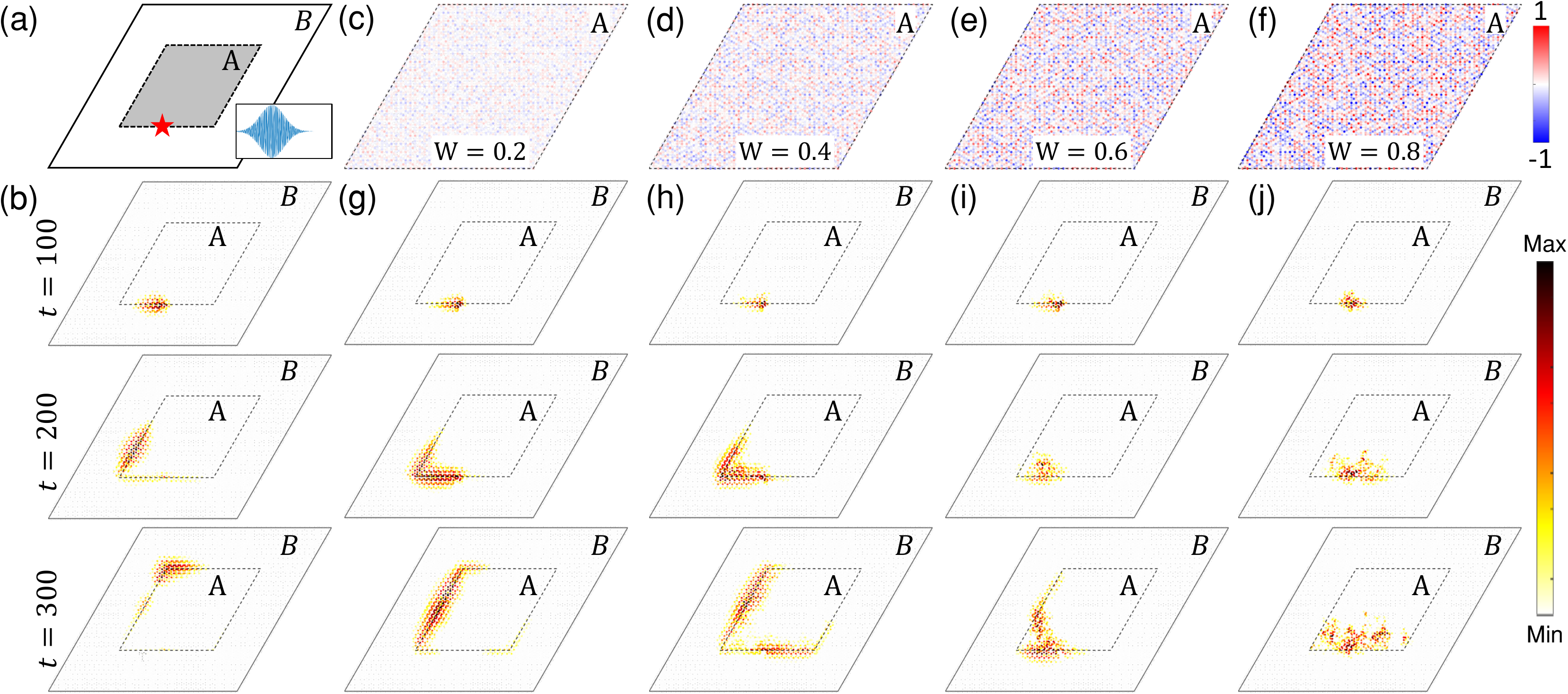}
\caption{(a) Numerical setup of the rhombic structure. The interface between two kinds of topologically distinct lattices is indicated by the black dashed line. (b) Transportation of the pseudospin-dependent wave package along the interface at a different time in a clean system. Color intensity represents the power spectral density of the particle displacements. (c)-(f) Schematics of the disordered inner core under random discrete disorder with strength $W=0.2$, $W=0.4$, $W=0.6$, and $W=0.8$ respectively. (g)-(j) Snapshots of transient simulated results under pseudospin excitation in the corresponding disordered structures. Colorbar represents the magnitude of the nondimensionalized displacement given as $20 \log{\frac{u_j}{A_0}} $, where $u_j$ is the particle displacement and $A_0$ is the amplitude of the displacement input.}
\label{C5FIG3}
\end{figure}

Figure~\ref{C5FIG3}(a) displays a schematic of a numerical simulation setup, which contains $40 \times 40$ unit cells in total and is made of two kinds of unit cells. Specifically, the inner core (indicated by the shaded region) consists of $20 \times 20$ topologically nontrivial unit cells (marked as type A in Fig.~\ref{C5FIG3}(a) for the sake of visualization) with $K_{in}=0.9$ and $K_{out}=1.2$. All the surrounding area contains trivial unit cells (type B) with $K_{in}=1.1$ and $K_{out}=0.8$. Therefore, a rhombic domain wall is formed between these two topologically distinct substructures. To selectively excite a pseudospin mode, we impose displacement input near the interface according to the pseudospin up state $p_{+}$. Specifically, Gaussian-modulated sinusoidal pulses (GMSP) with proper phases are placed at $6$ sites within a unit cell as indicated by the red star. Figure~\ref{C5FIG3}(b) shows snapshots of simulated elastic wave propagation under a pseudospin up excitation with center angular frequency at $1.71$ in a clean system. It is clear that the elastic waves travel only along the interface in the clockwise direction and can pass the sharp corners with negligible backscattering.

Then we move on to the disordered QSH systems to investigate their characteristics during the topological phase transitions. As we mentioned in the last section, introducing disorder in the trivial configuration will not flip the topological phase. We only impose disorder in the inner core region during the following analysis, which is made of topologically nontrivial unit cells [see the shaded area in Fig.~\ref{C5FIG3}(a)]. Figures~\ref{C5FIG3}(c)-(f) are the schematics of the inner core under discrete disorder with strength $W=0.2$, $W=0.4$, $W=0.6$, and $W=0.8$, respectively. Color intensity represents the disorder magnitude of spring stiffness ($W_j \epsilon_j$). 

Figures~\ref{C5FIG3}(g)-(j) show the snapshots of transient results of the elastic wave fields under a pseudospin up excitation with center frequency at $0.272$ Hz for corresponding disorder configuration. Similar to a clean system, the pseudospin waves mainly travel along the interface in the clockwise direction in a system under weak disorder $W = 0.2$ and $0.4$. However, as the $C_6$ symmetry at the interface is further broken by the disorder besides the initial mismatch between the topologically distinct unit cells, we start to observe more backscattering by the spin mixing defects~\cite{deng2017,chen2018,chaunsali2018} [See bottom panels of Figs.~\ref{C5FIG3}(g) and (h)]. When the stronger disorder is present, we can barely observe any propagation along the interface. Instead, we find the energy tends to penetrate into the bulk for $W=0.6$ (Fig.~\ref{C5FIG3}(i)) and show stronger localization for $W=0.8$ (Fig.~\ref{C5FIG3}(j)). This could be explained by the Anderson localization~\cite{anderson1958}. Based on the characteristic and significance of the propagation of pseudospin waves, we can learn that the inner core can retain its topological nature under weak disorder ($W < 0.4$) and start to lose its topological properties with the increase of disorder strength, which is inconsistent with our previous prediction by spin Bott index. 

It is worth mentioning that the spin Bott index of a disordered QSH system in Fig.~\ref{C5FIG2}(a) is an averaged result of numerous realizations, which actually shows a very large deviation during the transition region. The transient simulation results shown in Fig.~\ref{C5FIG3} are based on one particular realization, which does not reflect the statistical property of the topologically disordered QSH systems, particularly during the transition region. Please find the transient results of two more realizations under discrete disorder in Appendix E to see the disorder-realization-dependent propagation of pseudospin waves in disordered structures.

\section{Disordered topological phononic crystal}

\begin{figure}[h]
\includegraphics[width=.8\columnwidth]{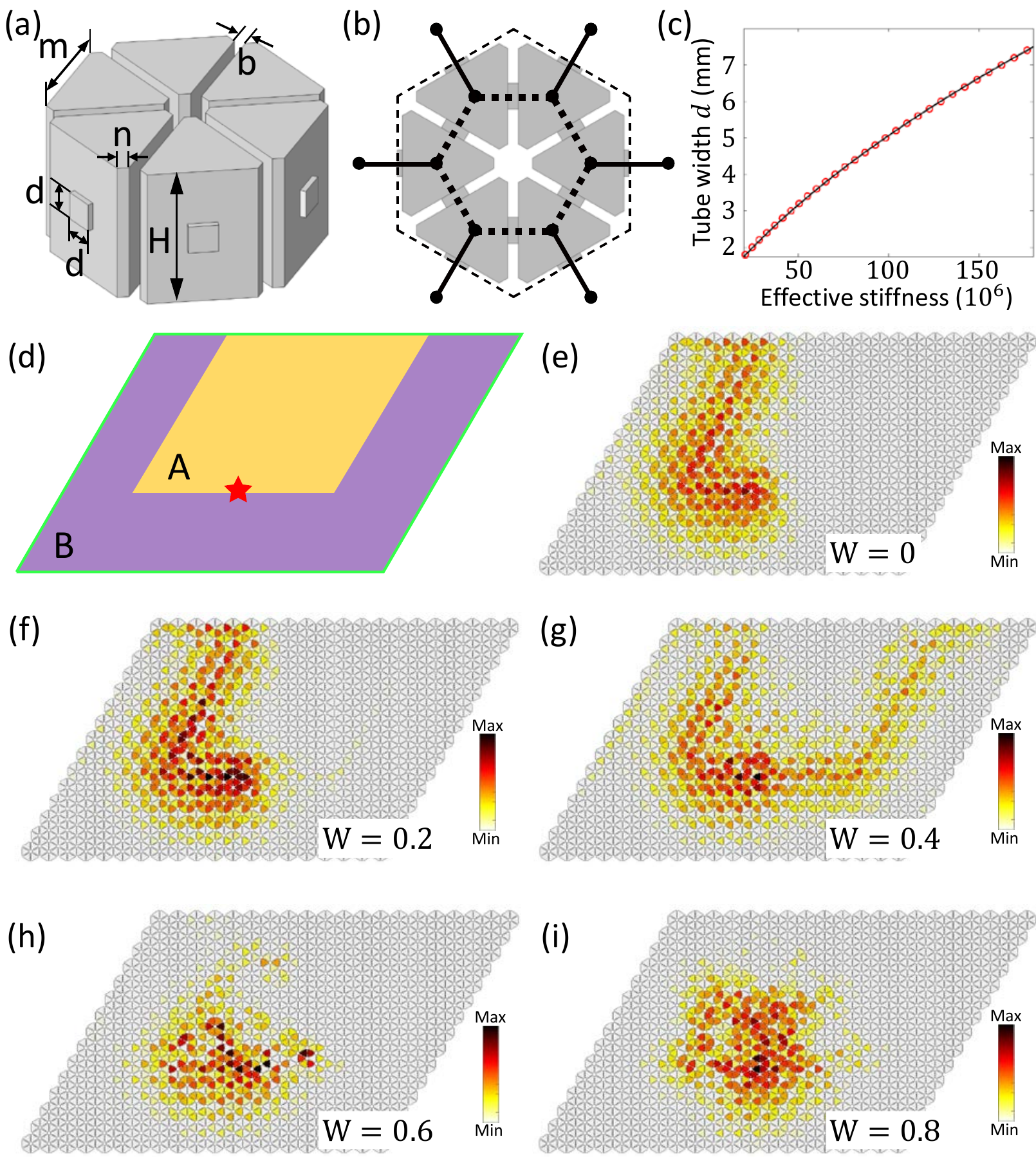}
\caption{(a) A detailed view of the design of the phononic crystal. (b) A top view of the unit cell and a mapping relation to the discrete model. (c) Relationship between the width $d$ of air tubes and the effective spring stiffness. (d) Schematic of the Finite element simulation setup. A U-shape interface is created by putting substructures with two types of unit cells adjacently. The red star indicates the excitation point. (e)-(i) Simulated wave field with a harmonic excitation along the interface under discrete disorder with disorder strength $W = 0$, $0.2$, $0.4$,$0.6$, and $0.8$, respectively.}
\label{C5FIG4}
\end{figure}

So far, the interplay between disorder and topology has been explored in the discrete model. To extend our discrete spring-mass model into a continuous structure, we propose a phononic crystal constructed by connecting hexagonal prism cavities with square tubes, as shown in Fig.~\ref{C5FIG4}(a). Here the gray parts are filled with air with density $\rho = 1.3$ $\mathrm{kg m^{-3}}$ and speed of sound $\nu = 343$ $\mathrm{m/s}$, and the white parts are rigid walls. The prism resonance cavities work effectively as the mass particles in the discrete model and the air channels in square windows are equivalent to the springs. The parameters used are $m=15$ $\mathrm{mm}$, $n=1.3$ $\mathrm{mm}$, $H=25$ $\mathrm{mm}$, and $b=2$ $\mathrm{mm}$. A top view of the extended unit cell of the phononic crystal and a rough mapping to the discrete spring-mass lattice can be seen in Fig.~\ref{C5FIG4}(b). By varying the cross-section of the air tubes, we can effectively tune the spring stiffness corresponding to the discrete lattice model. Figure~\ref{C5FIG4}(c) gives the relationship between the width of the \textit{square} channels ($d$) and the effective spring stiffness. Red circles represent the results obtained by fitting unit cell dispersion curves of a discrete lattice model with that of the phononic crystal calculated by finite element analysis (FEA) software COMSOL Multiphysics [See Appendix F for more details on the fitting process]. The black line stands for the second degree polynomial fitted to the data red dots. 

To verify the topological states along the interface between topologically nontrivial disordered QSH insulator and a trivial insulator, we construct a U-shape interface by joining the substructure with two kinds of unit cells. Specifically, type A unit cell is topologically nontrivial with $d_{in}=4.9$ $\mathrm{mm}$ and $d_{out}=5.4$ $\mathrm{mm}$, which corresponds to $K_{in}=0.9  \times 10^5 $ and $K_{out}=1.1 \times 10^5$ in the discrete model. Here, $d_{in}$ and $d_{out}$ refer to the width of intracell and intercell air channels, respectively. Type B unit cell is topologically trivial with $d_{in}=5.2$ $\mathrm{mm}$ and $d_{out}=4.7$ $\mathrm{mm}$, which is equivalent to a unit cell with $K_{in}=1.1 \times 10^5$ and $K_{out}=0.8 \times 10^5$ in the discrete model. Figure~\ref{C5FIG4}(d) is a schematic of the FEA setup. Specifically, the inner core consists of the type A unit cells with intracell air channels narrower than the intercell ones, while the outer part contains trivial unit cells (type B) with intracell air channels wider than the intercell ones. The green lines represent the radiative boundary conditions imposed on four sides of the system, which allows acoustic waves to leak into the environment to reduce reflection at these boundaries. Six point sources with appropriate phase differences are put in the six air cavities within one unit cell as marked by the red star in Fig.~\ref{C5FIG4}(d) to excite the system with pseudospin up modes. We then use COMSOL Multiphysics to perform the harmonic finite element analysis. Similar to our arrangement in the transient analysis in section~\ref{C5S4}, the discrete disorder is only introduced on the inner core region. Note that, the disorder configuration is determined by changing the air tube width $d$ based on the spring stiffness information from a disorder realization in the discrete model. See Appendix G for a flowchart explaining the process of constructing the disordered phononic structure. 

Figure~\ref{C5FIG4}(e) shows the steady state response of a clean system under pseudospin-up excitation. It is clear that the acoustic waves are very well confined to the topological interface and can only propagate in the clockwise direction. Particularly, the pseudospin waves can robustly pass the sharp bends without obvious reflections and scatterings. 

The one-way propagation of pseudospin dependent waves at $2.72$ kHz with the weak disorder ($W = 0.2, 0.4$) are shown in Figs.~\ref{C5FIG4}(f) and (g). Specifically, a disordered QSH system with A type unit cells under disorder level $W = 0.2$ or $0.4$ has spin Bott index 1 with very small variations, thus indicating a stable topologically nontrivial state [see Figure~\ref{C5FIG2}(a)]. In this case, we can observe that the acoustic waves are still mostly localized near the topological boundary, and a significant amount of the energy is still flowing in the clockwise direction. However, instead of a uniform energy distribution along the propagation path as seen in the clean system, we start to see more unevenly distributed modes at the interface with energy penetrating to the bulk of the substructure. 

As we further increase the disorder strength to $W=0.6$, the pseudospin excitation does not generate unidirectional propagating waves [see Fig.~\ref{C5FIG4}(h)]. Finally, when the disorder strength reaches $W=0.8$, the inner substructure becomes topological trivial with a spin Bott index very close to zero [see Fig.~\ref{C5FIG2}(a)]. That is, the interface between the substructures A and B can no longer be treated as a topological domain wall. As presented in Fig.~\ref{C5FIG4}(i), acoustic energy spreads into the bulk of the structure. Again the results in Figs.~\ref{C5FIG4}(e)-(i) are based on a single disorder realization, which cannot represent the statistical characteristics of the disordered topological QSH insulator. Steady-state wave fields under pseudospin up excitation for one more realization are plotted in Appendix H for comparison.

\section{Conclusion}
This paper investigates the interplay between topology and disorder in both discrete and continuous 2D phononic systems. Quantitative analysis of the disorder effect is conducted by tracking the spin Bott index. We find that a topologically nontrivial QSH system can endure a certain level of disorder and will eventually alter its topological nature under strong disorder. Transient simulation results of the pseudospin dependent waves also confirm the topological phase transition induced solely by disorder and agreed well with the spin Bott index prediction. With the proposed framework, we provide a powerful tool to quantitatively analyze the topological phases in a disordered 2D system. The results reveal the robustness of the zone-folding induced topological phases and may inspire future explorations on the sensitivity of the topologically nontrivial system to different types of disorders. 

While this study focused on the linear response of the 2D system, we envision that the findings of this study can be useful for the exploration of the nonlinear response of the 2D disordered topological system. Likewise, the extension of the spin Bott index to different types of systems (e.g., quasi-periodic 2D system or general 3D architecture) can be one of directions for the future studies. Lastly, the findings of this study may lay a foundation for the possible realization of TAI in the 2D/3D mechanical system.

\begin{acknowledgments}
We thank Dr. Terry Loring (University of New Mexico), Dr. Ying Wu (Nanjing University of Science and Technology), and Dr. Feng Li (Beijing Institute of Technology) for fruitful discussions. We also thank Dongxian Wang, Mingfei Wang, and Zhou Hu at the Beijing Institute of Technology for assistance in the numerical simulation. X. S. and J. Y. are grateful for the financial support from the U.S. National Science Foundation (EFRI-1741685). R.Z. acknowledges the support from the National Science Foundation of China (Project 11991033). J.Y. acknowledges the support from the SNU-IAMD and the Brain Pool Plus program funded by the Ministry of Science and ICT through the National Research Foundation of Korea (0420-20220160).

\end{acknowledgments}

\section*{APPENDIX A: Berry curvature and Spin Chern number}\label{C5APA}
\begin{figure}[ht]
\includegraphics[width=.6\columnwidth]{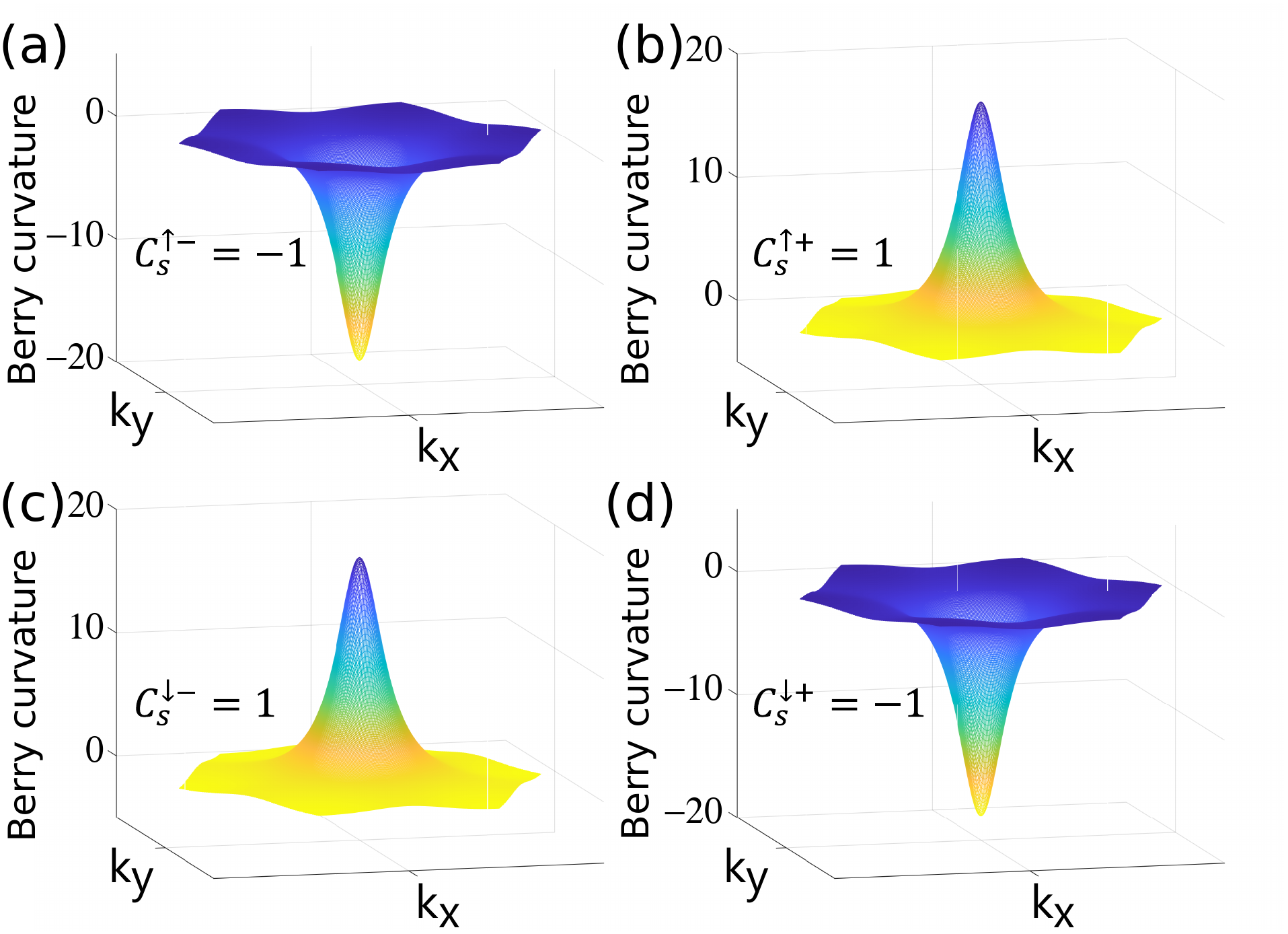}
\caption{Berry curvature and spin Chern number of four pseudo spin channels for $K_{in}<K_{out}$. (a) The lower band of pseudo spin up. (b) The upper band of pseudo spin up. (c) The lower band of pseudo spin down. (d) The upper band of pseudo spin down.}
\label{C5FIGSA}
\end{figure}

Figure~\ref{C5FIGSA} shows the calculated Berry curvature in the first Brillouin zone for all spin channels. By integrating the Berry curvature, we can obtain the corresponding spin Chern numbers of four spin bands, given as $C_{S}^{\downarrow-}=1$, $C_{S}^{\uparrow-}=-1$, $C_{S}^{\downarrow+}=-1$, and $C_{S}^{\uparrow+}=1$. Then the spin Chern numbers of spin up and spin down states are defined as $C_{S}^{\downarrow}=\frac{C_{S}^{\downarrow+}-C_{S}^{\downarrow-}}{2} = -1$ and $C_{S}^{\uparrow}=\frac{C_{S}^{\uparrow+}-C_{S}^{\uparrow-}}{2} = 1$, respectively. Therefore, the total spin Chern number can be calculated as $C_{S} = \frac{C_{S}^{\uparrow} - C_{S}^{\downarrow}}{2}$, indicating a topologically nontrivial system.

\section*{APPENDIX B: Dynamical matrix and effective Hamiltonian of finite size structure}\label{C5APB}
While assembling dynamical matrix $\mathbb{D}$, we need to arrange the nodal displacements such that particle displacements within a unit cell are packed in a subgroup with the increase of particle number as noted by the number in Fig.~\ref{C5FIG1}(a). Therefore, the modal displacement vector can be written as:
\begin{equation}
    \mathbb{U}_{6N \times 1}=
    \begin{bmatrix}
        \boldsymbol{U}_{1} \\
        \boldsymbol{U}_{2} \\
        \vdots \\
        \boldsymbol{U}_{N-1}  \\
        \boldsymbol{U}_{N} \\
    \end{bmatrix} = 
    \begin{bmatrix}
        \boldsymbol{T}_{s2n} \boldsymbol{\psi}_{1} \\
        \boldsymbol{T}_{s2n} \boldsymbol{\psi}_{2} \\
        \vdots \\
        \boldsymbol{T}_{s2n} \boldsymbol{\psi}_{N-1}  \\
        \boldsymbol{T}_{s2n} \boldsymbol{\psi}_{N} \\
    \end{bmatrix} = 
    \begin{bmatrix}
        \boldsymbol{T}_{s2n} & 0 & 0 & 0 & 0 \\
        0 & \boldsymbol{T}_{s2n} & 0 & 0 & 0 \\
        0 & 0 & \ddots & 0 & 0 \\
        0 & 0 & 0 & \boldsymbol{T}_{s2n} & 0 \\
        0 & 0 & 0 & 0 & \boldsymbol{T}_{s2n} \\
    \end{bmatrix}
    \begin{bmatrix}
        \boldsymbol{\psi}_{1} \\
        \boldsymbol{\psi}_{2} \\
        \vdots \\
        \boldsymbol{\psi}_{N-1}  \\
        \boldsymbol{\psi}_{N} \\    
    \end{bmatrix} = 
    \mathbb{T}_{6N \times 4N}
    \boldsymbol{\Psi}_{4N \times 1}
\end{equation}

\noindent where $N$ is the total number of unit cells, $\boldsymbol{U}_{j}$ is the modal displacement vector of the $j$th unit cell, $\boldsymbol{\psi}_{j}$ is the transformed modal displacement vector corresponding to the $j$th unit cell in the spin vector basis, $\mathbb{U}$ is the sorted modal displacements of the finite size structure, and $\mathbb{T}$ is the global transformation matrix. Then the effective Hamiltonian of the finite size structure is defined as: 
\begin{equation}
    \mathbb{H}_{4N \times 4N}=\mathbb{T}_{4N \times 6N}^{\dagger} \mathbb{D}_{6N \times 6N} \mathbb{T}_{6N \times 4N}.
\end{equation}

\section*{APPENDIX C: Eigenfrequencies and typical eigenmodes of the finite size structures with different types of unit cells}\label{C5APC}

\begin{figure}[ht]
\includegraphics[width=\columnwidth]{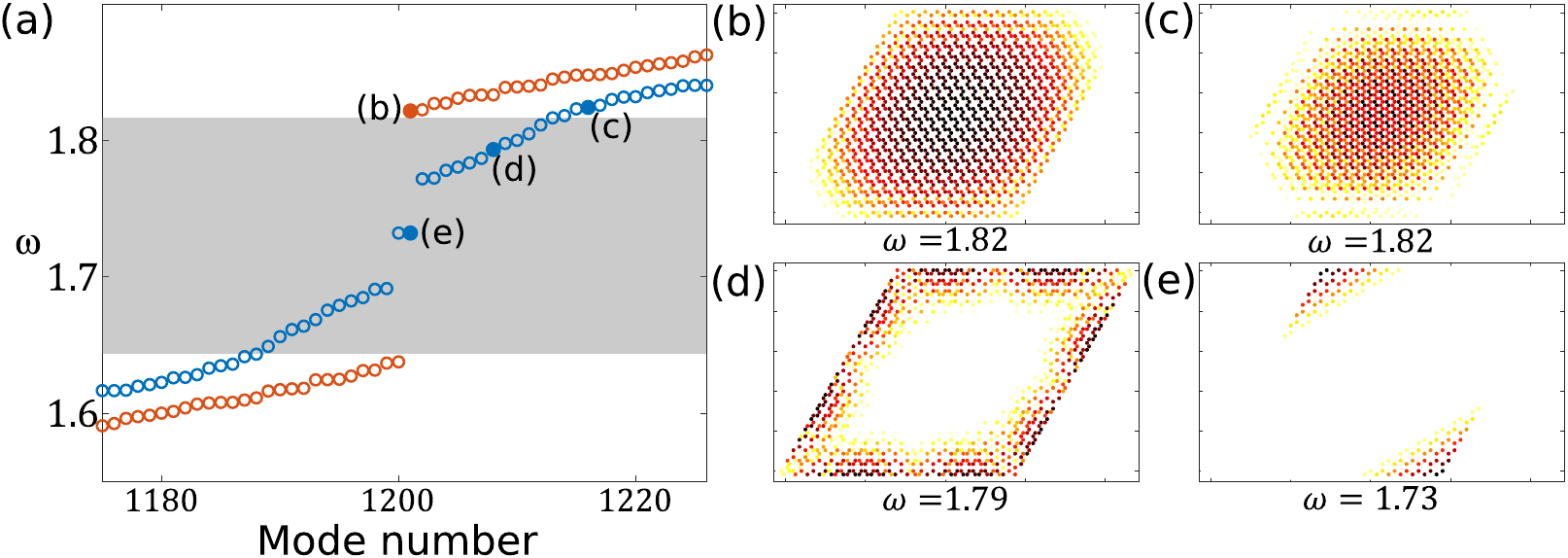}
\caption{(a) Eigenfrequency spectrum of the topologically trivial (red circles) or nontrivial (blue circles) structure. (b)-(e) Eigenmodes of the finite structure corresponding to the markers in (a).}
\label{C5FIGSB}
\end{figure}

Figure~\ref{C5FIGSB}(a) compares eigenfrequencies of the finite size structures with different choices of unit cells. The blue circles represent the case for topologically nontrivial unit cells with $K_{in}=0.9 < K_{out}=1.2$, which is the same as Fig.~\ref{C5FIG1}(e). The red circles stand for the results of topologically trivial unit cells with $K_{in}=1.1 > K_{out}=0.8$. Figures~\ref{C5FIGSB}(b) and (c) give examples of bulk modes with frequencies outside the band gap in a trivial and nontrivial configuration, respectively. While the properties of the bulk modes are similar, clear differences can be observed in the frequency band gap It is easy to catch that the main difference between these two cases is the emergence of boundary modes within the band gap range in the topologically nontrivial structure. Figures~\ref{C5FIGSB}(d) and (e) show typical examples of an edge state and a corner state, respectively.

\section*{APPENDIX D: Effect of disorder parameters on the topological phase transitions }\label{C5APD}

\begin{figure}[ht]
\includegraphics[width=.8\columnwidth]{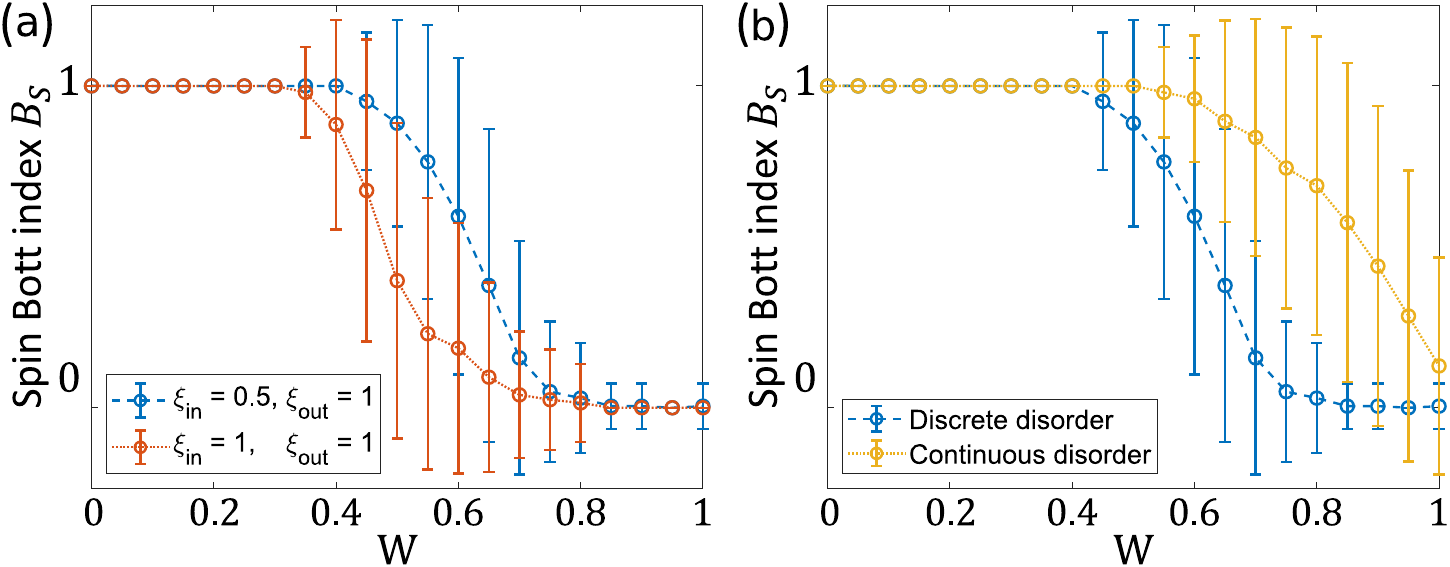}
\caption{Effect of (a) disorder ratio  and (b) disorder type on the topological phase transition.}
\label{C5FIGSC}
\end{figure}

In the main body of this paper, we focus on the interplay of discrete disorder with topology. It is controlled by two parameters $\xi_j$ and $\epsilon_j$. Specifically, $\xi_j$ refers to the disorder ratio between intercell and intracell disorder coefficients ($\xi_{in}/\xi_{out}$). It explains the fact that intercell and intracell springs may have different tendencies to develop disorders. Figure~\ref{C5FIGSC}(a) compares the disorder averaged spin Bott index for different combinations of $\xi_{in}$ and $\xi_{out}$. Blue line stands for the results of the case with $\xi_{in}=0.5$, $\xi_{out}=1$, which is the same as Figure~\ref{C5FIG2}(a). Red line plots the results obtained with $\xi_{in} = 1$, $\xi_{out}=1$. The choice of $\epsilon_j$ determines the disorder type. By randomly selecting \textit{integers} from the discrete uniform distribution on the interval $[-1,1]$, we define the so-called discrete disorder, which is a very simplified case of disorder. A more general type of disorder, namely the continuous disorder, can be realized by setting $\epsilon_j$ a uniformly distributed number in the interval $[-1,1]$. Figure~\ref{C5FIGSC}(b) compares the effects of discrete (blue line) and continuous (yellow line) disorder on topology. By varying disorder parameters, we will be able to manipulate the topological phase transition process and shift the critical transition boundary.

\section*{APPENDIX E: Transient wave analysis for other realizations under discrete disorder}\label{C5APE}
\begin{figure}[ht]
\includegraphics[width=\columnwidth]{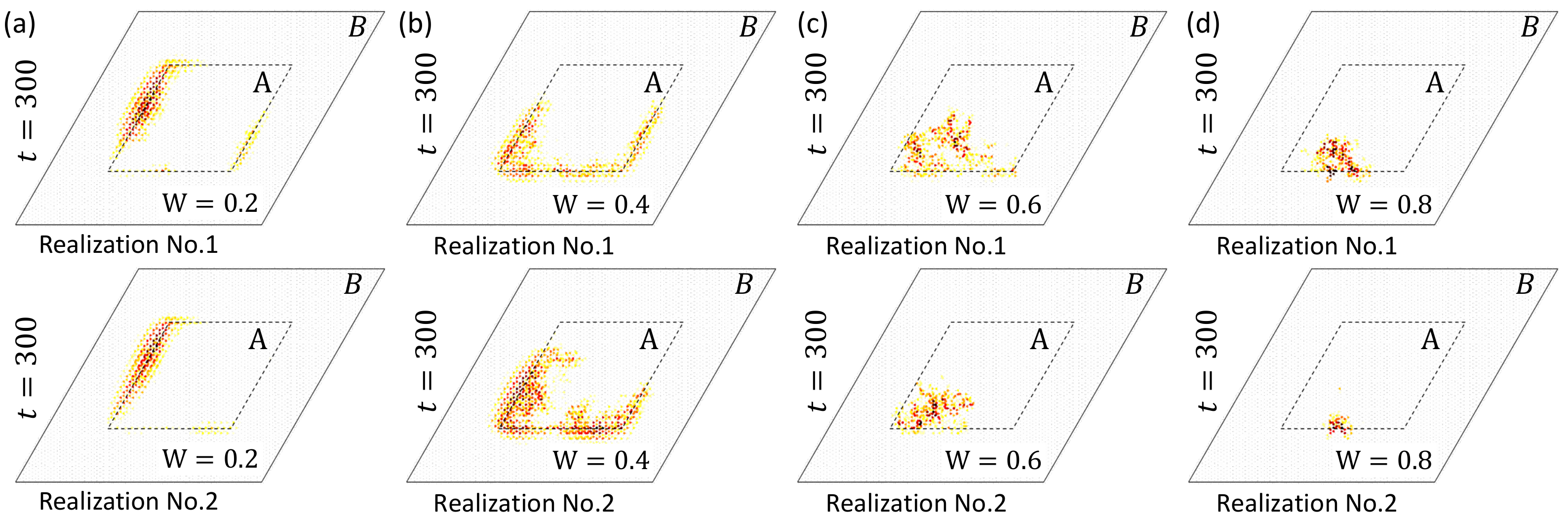}
\caption{(a)-(d) Transient simulated wave fields at $t=300$ under pseudospin excitation for two more realizations under random discrete disorder with disorder strength $W=0.2$, $W=0.4$, $W=0.6$, and $W=0.8$, respectively.}
\label{C5FIGSD}
\end{figure}

In Fig.~\ref{C5FIG2}, we can observe that the spin Bott index of a system with a certain disorder level shows a very large deviation, which indicates that topology is highly dependent on a particular disorder configuration for a certain range of disorders. Here, we conduct $2$ extra transient simulations for different discrete disorder realizations and demonstrate the snapshots of wave fields at $t=300$ in Fig.~\ref{C5FIGSD}. We see very similar wave propagation phenomena for weak disorder cases ($W=0.2$ and $W=0.4$), corresponding a the stable spin Bott index that is very well quantized at $1$. Figures~\ref{C5FIGSD}(c)-(d) show two more numerical simulation results with disorder strength $W=0.6$ and $0.8$, respectively. By comparing the structure responses of different realizations, we find that a larger standard deviation of the spin Bott index suggests that systems with different disorder realizations tend to respond more distinctively under the same pseudospin excitation. Despite the fact such an approximate relationship involves intuitive interpretations of the transient wave fields, it still helps to give a better understanding of the disordered QSH system based on the calculation of the spin Bott index.

\section*{APPENDIX F: Parameter fitting between the square tube width and the effective spring stiffness} \label{C5APF}
\begin{figure}[ht]
\includegraphics[width=.5\columnwidth]{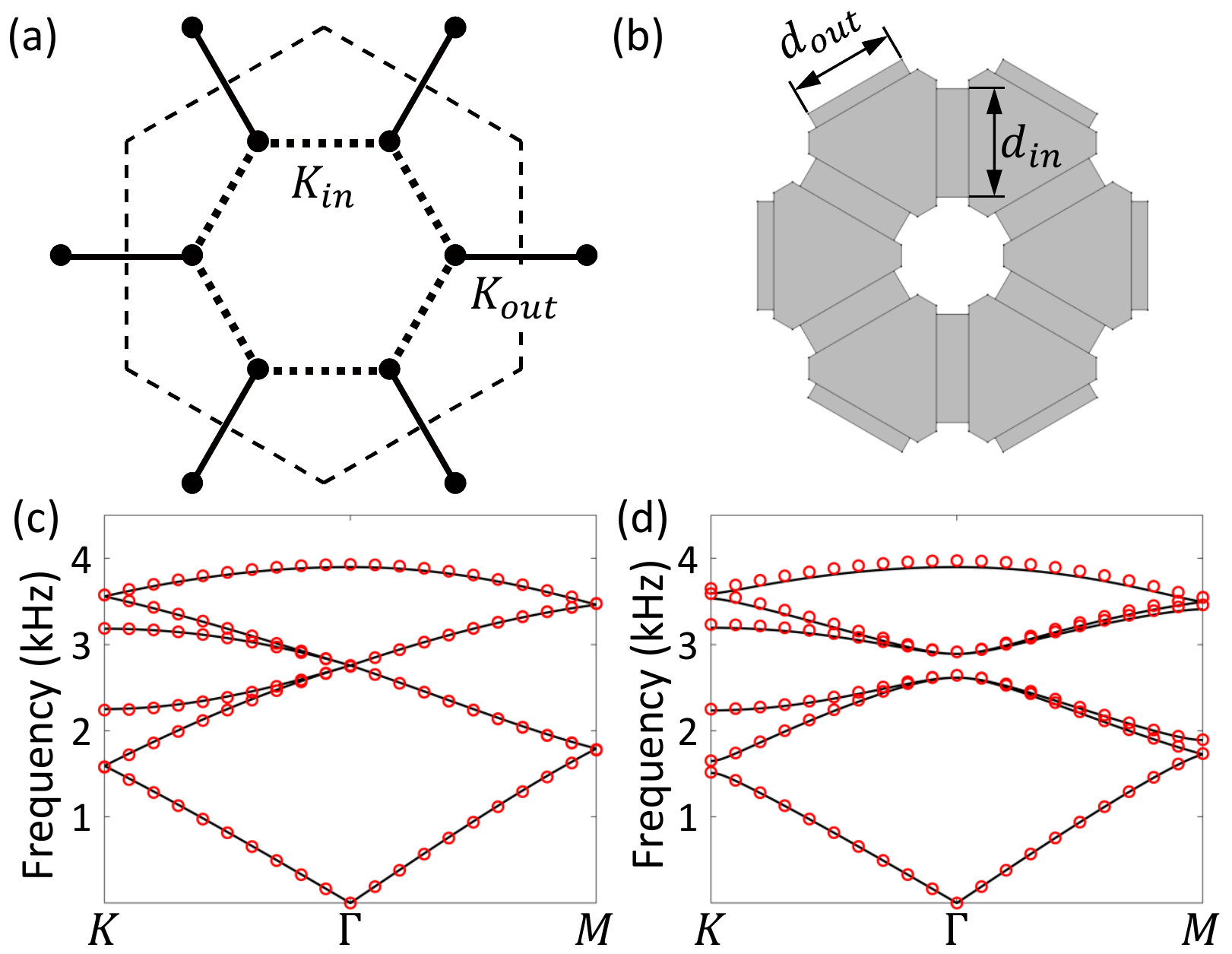}
\caption{(a) Unit cell of the discrete model (b) Top view of the unit cell of a phononic crystal. (c)-(d) Comparison of unit-cell dispersion curves with a double Dirac point or a band gap obtained from the spring-mass model (solid lines) and the phononic crystal (red dots)}
\label{C5FIGSE}
\end{figure}

In order to find the relationship between the square tube width and the effective spring stiffness, we fit the dispersion diagram of the spring-mass model with that of a phononic crystal obtained from COMSOL Multiphysics. Figures~\ref{C5FIGSE}(a)-(b) show the unit cell of the spring-mass model and the phononic crystal, respectively. A comparison between the dispersion relations obtained from the discrete model and the continuous model is plotted in Fig.~\ref{C5FIGSE}(c). Black lines represent the result of a discrete spring-mass model with $K_{in}=K_{out}=1.0 \times 10^8$. Red dots stand for the FEA result with $d_{in}=4.73$ mm, and $d_{out}=5.75$ mm. Figure~\ref{C5FIGSE}(d) shows a similar results for $K_{in}=0.9 \times 10^8$, $K_{out}=1.2 \times 10^8$ and $d_{in}=d_{out}=5.08$ mm. By varying $d_{in}$ and $d_{out}$ values and repeating the dispersion curves fitting process, we finally come up with the relationship between the square tube width $d$ and the effective spring stiffness $K$ within the range of our interest, as shown in Fig.~\ref{C5FIG4}(c).

\section*{APPENDIX G: Construction of disordered phononic crystal structure}\label{C5APG}

\begin{figure}[ht]
\includegraphics[width=.5\columnwidth]{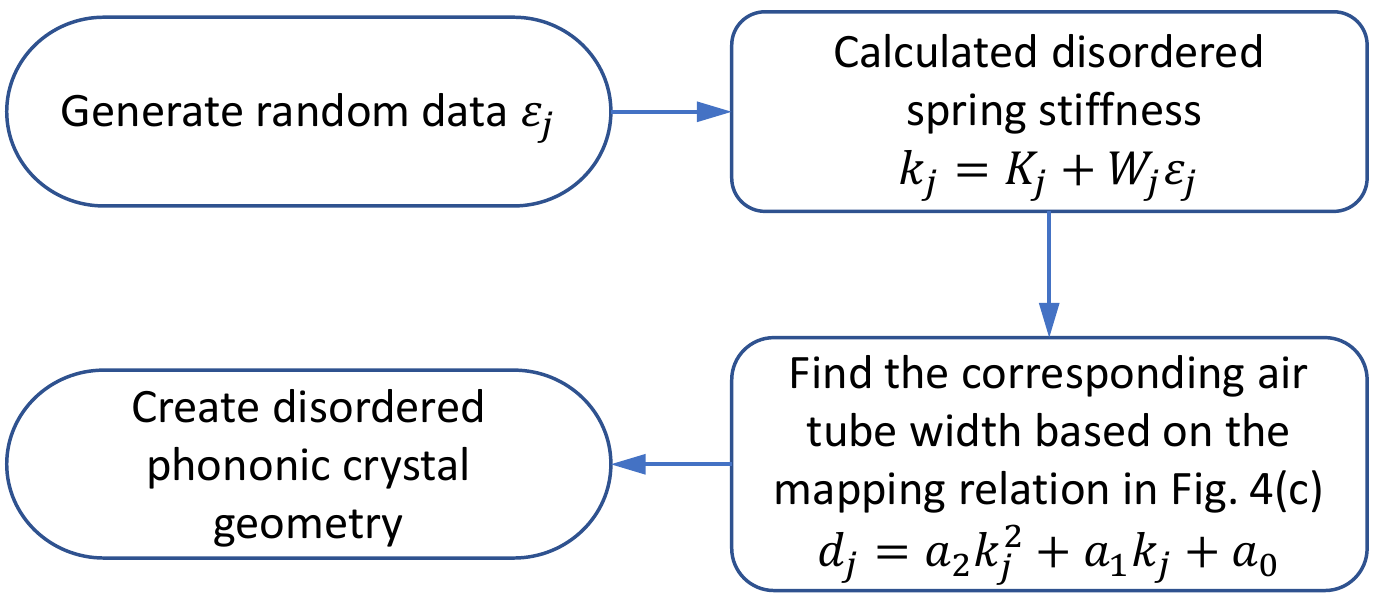}
\caption{Procedure to generate the disordered phononic crystal.}
\label{C5FIGSF}
\end{figure}
The disordered phononic structures are constructed by mapping from discrete spring-mass systems. First, we need to generate the random parameter $\epsilon_j$ that determines the disorder configuration, where $j$ is the spring element number. Then we substitute $\epsilon_{j}$ into Eq.~\ref{C5eq:5} to calculate the stiffness of the $j$th disordered spring. Next, based on the fitted polynomial in Fig.~\ref{C5FIG4}(c), we find the width $d$ of an air tube corresponding to each spring in the discrete model. Finally, we draw the geometry of disordered phononic crystal for FEA simulation in COMSOL Multiphysics. Figure~\ref{C5FIGSF} summarizes the procedure to generate the disordered phononic crystal model. 

\section*{APPENDIX H: Steady state in phononic crystal for extra discrete disorder realization}\label{C5APH}

\begin{figure}[ht]
\includegraphics[width=.8\columnwidth]{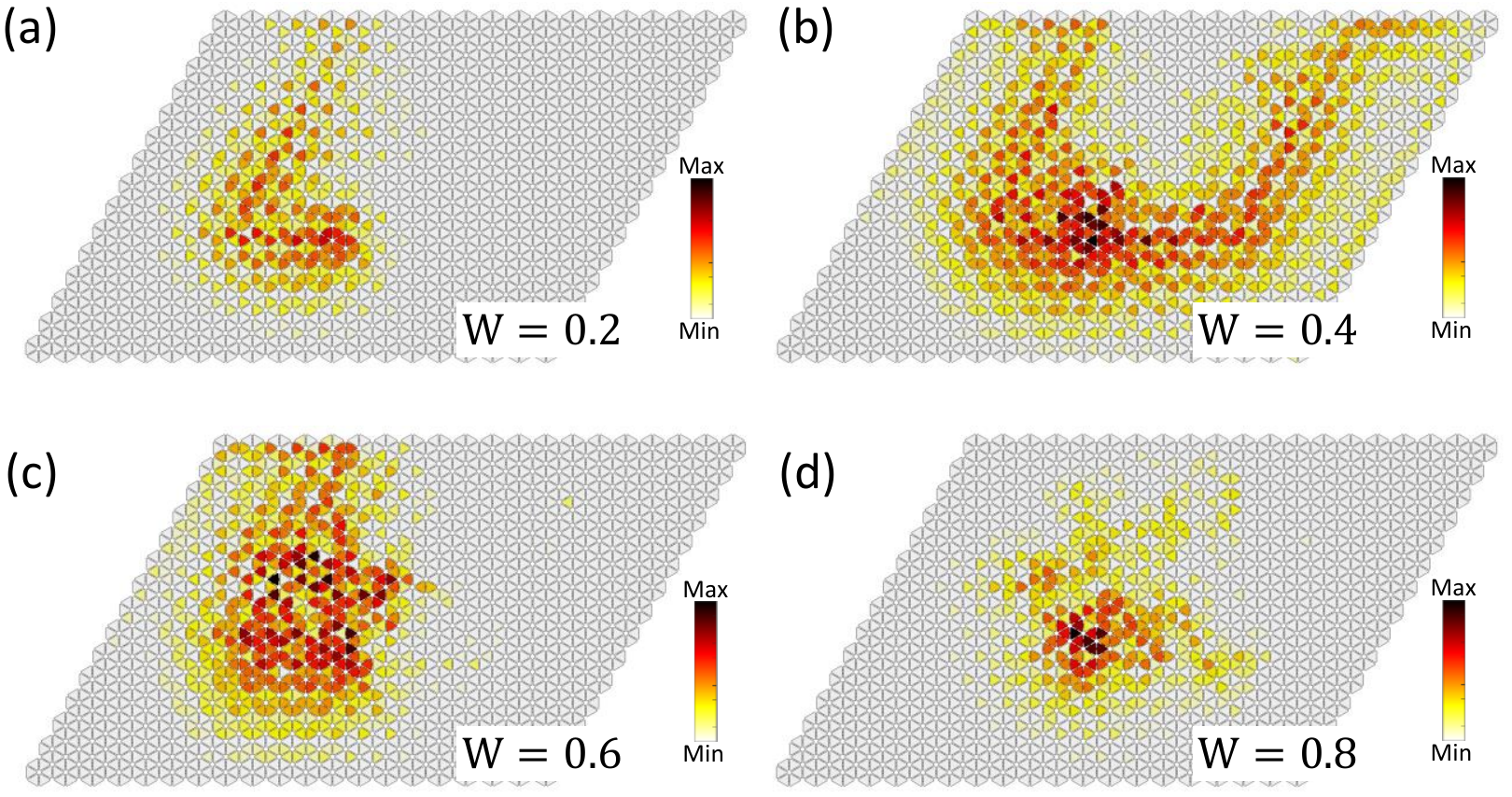}
\caption{(a)-(d) Steady state wave field under a harmonic excitation in the presence of discrete disorder with disorder strength $W = 0.2$, $0.4$, $0.6$, and $0.8$, respectively.}
\label{C5FIGSG}
\end{figure}

In Figure~\ref{C5FIGSG}, we show the steady state of disordered phononic crystal under pseudospin excitation with the same disorder strength but a different configuration compared to realization in Figure~\ref{C5FIG4}. As we can see, with the increase of disorder strength, less energy could stay confined near the interface and propagate in the clockwise direction. We observe a growing leakage into the bulk and more significant energy localization. In general, by comparing the results in Fig.~\ref{C5FIG4} and Fig.~\ref{C5FIGSG}, we can know that wave propagates in a very similar trend for disordered systems with spin Bott index of low deviation ($W=0.2$, $0.4$, and $0.8$). When the averaged spin Bott index has a large variation ($W=0.6$), a more drastic difference may appear in the disordered system.



\bibliography{2D_Disorder} 
\bibliographystyle{apsrev4-2}

\end{document}